\documentclass[prb,showpacs,twocolumn]{revtex4}

\usepackage{graphicx}
\usepackage{dcolumn}
\usepackage{bm, array}
\usepackage{multirow}%
\usepackage[dvipsnames]{xcolor}
\usepackage{longtable}
\usepackage{latexsym}

\def\uc{\texttt{\footnotesize uc}}
\def\at{\texttt{\footnotesize at}}
\def\sc{\texttt{\footnotesize sc}}
\def\mix{\texttt{\footnotesize mix}}
\def\gen{\texttt{\footnotesize gen}}
\def\pure{\texttt{\footnotesize pure}}
\def\sol{\texttt{\footnotesize sol}}
\def\ssc{\texttt{\tiny sc}}
\def\suc{\texttt{\tiny uc}}
\def\mix{\texttt{\footnotesize mix}}

\def\at{\texttt{\footnotesize at}}
\def\bin{\texttt{\footnotesize bin}}
\def\A{\texttt{\footnotesize A}}
\def\B{\texttt{\footnotesize B}}
\def\sA{\texttt{\footnotesize A}}
\def\sB{\texttt{\footnotesize B}}
\def\sS{\texttt{\tiny S}}
\def\sI{\texttt{\tiny I}}
\def\ssA{\texttt{\tiny A}}
\def\ssB{\texttt{\tiny B}}
\def\Mg{\texttt{\footnotesize Mg}}

\def\omu{\overline{\mu}}

\begin{document}
\textheight 226.5mm

\title{First-principle solubilities of alkali and alkaline earth metals in Mg-B alloys}

\author{Roman V. Chepulskii and Stefano Curtarolo}

\affiliation{Department of Mechanical Engineering and Materials Science, and Center for Theoretical and Mathematical Sciences, Duke University, Durham, North Carolina 27708, USA}

\date{\today}

\begin{abstract}
  By devising a novel framework, we present a comprehensive theoretical study of solubilities of alkali (Li, Na, K, Rb, Cs)
  and alkaline earth (Be, Ca, Sr, Ba) metals in the he boron-rich Mg-B system.
  The study is based on first-principle calculations of solutes formation energies in MgB$_2$, MgB$_4$, MgB$_7$ alloys
  and subsequent statistical-thermodynamical evaluation of solubilities. The advantage of the approach consists in considering
  all the known phase boundaries in the ternary phase diagram.
  Substitutional Na, Ca, and Li demonstrate the largest solubilities, and Na has the highest (0.5-1 \% in MgB$_7$ at $T=650-1000$ K).
  All the considered interstitials have negligible solubilities.
  The solubility of Be in MgB$_7$ can not be determined because the corresponding low-solubility formation
  energy is negative indicating the existence of an unknown ternary ground state.
  We have performed a high-throughput search of ground states in binary Mg-B, Mg-$A$, and B-$A$ systems,
  and we construct the ternary phase diagrams of Mg-B-$A$ alloys based on the stable binary phases.
  Despite its high temperature observations, we find that Sr$_{9}$Mg$_{38}$ is not a low-temperature equilibrium structure.
  We also determine two new possible ground states CaB$_{4}$ and RbB$_{4}$, not yet observed experimentally.
\end{abstract}

\pacs{74.70.Ad, 61.50.Ks, 81.30.Hd}

\maketitle

\section{Introduction}
The interest in magnesium diboride emerged
after the discovery of superconductivity in MgB$_2$ at about $T_c$=39 K.\cite{MgB2-2001}
Attempts to increase $T_c$ by small additions of alkali
(Li\cite{Owens01Li,Zhao01Li,Zhang02CaLi,Cimb02Li,Karp08Li}, Na\cite{Toul02NaCa,Agost07Na}, Rb\cite{Palnich07RbCsBa,Singh08RbCs},
Cs\cite{Palnich07RbCsBa,Singh08RbCs}) and alkaline earth (Be\cite{Feln01Be}, Ca\cite{Toul02NaCa,Zhang02CaLi,Cheng03Ca,Tamp03CaSr},
Ba\cite{Palnich07RbCsBa}) metals to MgB$_2$,
proved to be unsuccessful.
The difficulty was attributed not only to the inability of such solutes to decrease $T_c$
but also to their low solubility and precipitation in secondary phases.
Although a claimed superconductivity at 50K was reported
for the Mg-B-$A$ ($A$=Cs, Rb, Ba) system \cite{Palnich07RbCsBa},
attempts to reproduce the results have been so far unsuccessful (for $A$=Cs, Rb) \cite{Singh08RbCs}.
The problem can be attributed to solutes' segregation in grain boundaries and
to thus to low solubility in the bulk phase (much lower than reported in Ref. \onlinecite{Palnich07RbCsBa}).

For consistent interpretation of experimental observations,
theoretical studies of solubility in Mg-B of alkali metals and
alkaline earth metal solubilities are therefore necessary.
For instance, in Ref. \onlinecite{WuGao05}, the semi-empirical Miedema
approach \cite{Miedema80}and the Toop's model \cite{Toop65} were used to address the
the heats of formation of binary alloys Mg-$A$, B-$A$, and Mg-B ($A$=Li, Na, and Ca).
It was proposed that Ca may form stable compounds while Na, Li could lead to meta-stable or unstable
ternary phases in MgB$_2$.
In Ref. \onlinecite{BernMass06}, by
calculating first-principle formation energies of Li and Na
impurities in MgB$_2$, and by neglecting the effects of other ground states in the ternary phase diagram solubility calculations,
it was concluded that Na should have very low solubility, whereas the solubility of Li should be comparatively
higher although modestly diminished by a segregation into LiB phase.

The present paper is orthogonal to previous studied. We develop a comprehensive theoretical framework
to determine the solubilities of alkali metals (Li, Na, K, Rb, Cs) and alkaline earths (Be, Ca, Sr, Ba)
in the boron-rich Mg-B system.
The study consists of first-principle calculations of solutes formation energies in
MgB$_2$, MgB$_4$, MgB$_7$ alloys and subsequent statistical-thermodynamical evaluation
of solubilities with respect to {\it all known equilibrium states} of the Mg-B-$A$ system.
The results help outlining future directions in experimental searches.

The paper is organized as following. In Section \ref{SEC:Cages}, we
describe the adopted solubility mechanisms in Mg-B.
In Section \ref{SEC:FormEn}, we introduce the relevant impurity formation energies
in terms of supercell energy calculations and the appropriate ground state(s).
In Section
\ref{SEC:F1}, the approximation for the free energy of Mg-B-$A$
solid solution is formulated. In Section
\ref{SEC:solubility.theory}, we present an approach  for solubility
calculation considering all the ternary ground states. A simple
analytical low-solubility approximation is devised. Section
\ref{SEC:1pr} is devoted to the high-throughput {\it ab initio}
search for ground states in binary Mg-B, Mg-$A$, and B-$A$ systems,
to the ternary phase diagrams of Mg-B-$A$ systems,
and to the impurity formation energies determined through the phase boundaries of the systems.
The numerical values of solubilities are presented in Section \ref{SEC:solubility.results}.
Section \ref{SEC:conclusions}
summarizes the results, draws conclusions, and comments on
strategies for future research in this area.

\section{Mg-B-$A$ solid solution}
\label{SEC:Cages}

For the description of solubility of alkali and
alkaline earth metal elements $A$ ($A$=Li, Be, Na, Mg, K, Ca, Rb, Sr,
Cs, Ba) in the Mg-B system, we consider the disordered solid solutions
of $A$-atoms as interstitial and magnesium-substitutional impurities
in the experimentally reported compounds \cite{MgB-exper,Pauling}
MgB$_2$, MgB$_4$, and MgB$_7$.
We do not consider boron substitutions by $A$ because they has not been observed
experimentally\cite{Owens01Li,Zhao01Li,Zhang02CaLi,Cimb02Li,Karp08Li,Toul02NaCa,Agost07Na,Palnich07RbCsBa,Singh08RbCs,Feln01Be,Cheng03Ca,Tamp03CaSr}.
The disordered solid solution of $A$ inside Mg-B is labeled as
``(1)'' throughout the paper.

While the magnesium-substitutional positions are determined \cite{Pauling},
the ``most accommodating'' interstitial locations have to be found.
The task is implemented with the following exhaustive search performed
through our software {\small AFLOW}\cite{SC20,AFLOW}. Let us consider a
quadruplet of no-coplanar atoms, where the first atom belongs to
the unit cell and the others
are closer than the maximum diagonal of the unit cell to the first atom.
A {\it cage} is defined when the spherical region of space
touching all the four atoms of the quadruplet does not contain further atoms inside.
An {\it interstitial} position is found if the cage has its center inside the unit cell.
By considering all the possible combinations,
the symmetrically inequivalent interstitials can be identified through the
calculation of their site symmetry (with the factor group of the unit cell).
Note that in unit cells with complex arrangements,
many of the interstitials positions can be extremely close. Thus,
an interstitial atom located in any of those close positions
would deform the nearby local atomic environment and {\it relax} to the same final location.
Hence, the number of symmetrically inequivalent
cages can be further reduced by considering
the whole set of positions that would {\it agglomerate} upon insertion of an interstitial atom,
as a single interstitial position.
The results of the search are presented in Table \ref{Tab:Cages}.
The table demonstrates that the higher boron contents the larger number of cages
and the larger radius of the bigger cage.

\begin{table}[htb]
  \caption{Centers and sizes of the symmetrically inequivalent interstitial
    cages for the MgB$_2$, MgB$_4$, and MgB$_7$ unit cells.
    Coordinates are presented as fractions of the
    standard unit cell basis vectors $(a,b,c)$ \cite{Pauling}.
    The radius is the distance from the cage center to the nearest Mg or B atom.
    $\nu_i$ indicates the numbers of the symmetrically equivalent cages within the
    unit cell. The total numbers of distinct geometrical cages before agglomeration upon
    insertion of an interstitial is reported in brackets.} \label{Tab:Cages} \small
\begin{tabular}{c|c|c|c|c} \hline\hline
Structure & $i$ &Coordinates & Cage & $\nu_i$\\
         &     & (fract. of $a$,$b$,$c$)         & radius (${\AA}$) &  \\
\hline
MgB$_2$ & 1 & 0.341, 0.681,  0.000  &  1.7619 &   2 (6)\\
        & 2 & 0.006, 0.000,  0.500  &  1.7616 &   1 (6)\\
        & 3 & 0.500,  0.500,  0.122  &  1.6015 &   6 (6)\\
\hline
MgB$_4$ & 1 & 0.106, 0.669,  0.412  &  1.9145 &   4 (52)\\
(Mg$_4$B$_{16}$) & 2 &  0.375, 0.250,  0.337  &  1.7836 &   4 (4) \\
        & 3 & 0.145, 0.632,  0.813  &  1.7821 &   8 (24) \\
        & 4 & 0.144, 0.750,  0.925  &  1.7252 &   4 (24)\\
        & 5 & 0.225, 0.750,  0.719  &  1.6322 &   4 (4)\\
        & 6 & 0.086, 0.015,  0.527  &  1.6128 &   8 (8)\\
        & 7 & 0.101, 0.543,  0.261  &  1.5322 &   8 (8)\\
\hline
MgB$_7$ & 1 & 0.000, 0.049,  0.618  &  2.0050 &   8 (152)\\
(Mg$_8$B$_{56}$)& 2 &  0.189,  0.250  0.821   &  1.8937 &   8 (24)\\
        & 3 & 0.250,  0.217,  0.750   &  1.7995 &   8 (8)\\
        & 4 & 0.243,  0.250,  0.505   &  1.7888 &   8 (24)\\
        & 5 & 0.197,  0.250,  0.469   &  1.7802 &   8 (24)\\
        & 6 & 0.000,  0.250,  0.150   &  1.7002 &   4 (24)\\
        & 7 & 0.012,  0.000,  0.000   &  1.6995 &   8 (64)\\
        & 8 & 0.250,  0.062,  0.250   &  1.6068 &   8 (40)\\
        & 9 & 0.000,  0.171,  0.788   &  1.4991 &   8 (8)\\
        & 10 & 0.131, 0.186,  0.119   &  1.4443 &   16(16)\\
\hline\hline \end{tabular} \end{table}

\section{Formation energies definitions}
\label{SEC:FormEn}


\vspace{-3mm}
\subsection*{``Raw'' formation energies}
\vspace{-3mm}
Let us define the so-called ``raw'' formation energies $\mu_i$ and
$\omu_s$ (composition unpreserving \cite{Mishin})
as the changes of the energies of the solvent upon introduction of
one solute atom $A$ in the $i$-th type interstitial or $s$-th type substitutional positions.
In first-principles calculations, the solvent can
be replaced by a large supercell ({\it ``sc''}), so that
\begin{equation}
  \label{Eq:mIII} \begin{array}{c} \mu_i \equiv \mu_i^{(1)}(\A)=
    E_\sc[\A^{(i)}\Mg_{n_\ssc}\B_{m_\ssc}]-
    E_\sc[\Mg_{n_\ssc}\B_{m_\ssc}], \\\\
    \omu_s \equiv \mu_s^{(1)}(\A)-\mu^{(1)}(\Mg)= \\
    E_\sc[\A^{(s)}\Mg_{n_\ssc-1}\B_{m_\ssc}]-
    E_\sc[\Mg_{n_\ssc}\B_{m_\ssc}],
  \end{array}
\end{equation}
where $n_\ssc$ and $m_\ssc$ are the numbers of Mg and B atoms in the
supercell; $\mu^{(1)}(\Mg)$, $\mu_i^{(1)}(\A)$, and
$\mu_s^{(1)}(\A)$ are the zero-temperature chemical potentials of the Mg-atom,
and the $A$-atoms in the $i$-th interstitial and $s$-th substitutional positions, respectively \cite{note1}.

The conventional unit cells of MgB$_2$, MgB$_4$, and MgB$_7$ are
used to construct the appropriate supercells.
The supercells are chosen to have solute-solute
$A$-$A$ distance at least $\sim$2.5 times the nearest neighbor
solvent bonds in order to diminish the contribution of the $A$-$A$
interactions to the calculated energies.
We use $2\times2\times2$ unit cells for MgB$_2$, while there is no need to
create supercells of MgB$_4$ and MgB$_7$ since their unit cells are already large enough.
Note that boron-boron (B-B)
bond is the shortest among the Mg-Mg, B-B, and Mg-B bonds.
Table \ref{Tab:nm} summarizes
the parameters of the supercells
and Table \ref{TAB:CscIII} lists the supercell atomic compositions.

\begin{table}[htb]
  \caption{
    Numbers of atoms $n,m$ (Mg$_n$B$_m$)
    in the conventional unit cells ({\it ``uc''}) \cite{Pauling}
    and in the constructed supercells ({\it ``sc''}) for MgB$_2$, MgB$_4$, and MgB$_7$.
    $\nu_s$ indicates the number of symmetrically equivalent $s$-type Mg atom positions within the
    corresponding unit cell:
    $\sum_{s} \nu_{s} =n_\uc$.
    $d_{\sA\sA}$ and $d_{\sB\sB}$ are the distances between the nearest solutes and nearest solvent atoms
    (boron-boron in our case) in the supercell.}
  \label{Tab:CagesNN}\label{Tab:nm}\small
  \begin{tabular}{c|cc|c|cc|c|c} \hline\hline
    Compound & $n_{\uc}$ & $m_{\uc}$ & $\nu_s$ &  $n_\sc$
    & $m_\sc$ & $d_{\sB\sB}$ (${\AA}$) & $d_{\sA\sA}/d_{\sB\sB}$ \\
    \hline
    MgB$_2$  & 1 & 2  & 1   & 8  & 16  &  1.7811 & 3.46 \\
    MgB$_4$  & 4 & 16 & 4   & 4  & 16  &  1.7004 & 2.60 \\
    MgB$_7$  & 8 & 56 & 4,4 & 8  & 56  &  1.7372 & 3.44  \\
    \hline\hline
  \end{tabular}
\end{table}
\begin{table}[htb]
  \caption{Interstitial and substitutional supercell
    atomic compositions
    ($x_\A^\sc$,$x_\B^\sc$,$x_\Mg^\sc$)
    as functions of the $n_\sc$ (for Mg) and $m_\sc$ (for B)
    numbers of atoms in the super cell.}\label{TAB:CscIII}
  \begin{tabular}{c|c|c} \hline\hline
    &  Interstitials  & Substitutional \\ \hline
    {\footnotesize $x_\A^\sc$} &
    $[1+n_\sc+m_\sc]^{-1}$&$[n_\sc+m_\sc]^{-1}$ \\
    {\footnotesize $x_\B^\sc$}&$m_\sc[1+n_\sc+m_\sc]^{-1}$&
    $m_\sc[n_\sc+m_\sc]^{-1}$ \\
    {\footnotesize $x_\Mg^\sc$}&$n_\sc[1+n_\sc+m_\sc]^{-1}$&
    $n_\sc[n_\sc+m_\sc]^{-1}$ \\
    \hline\hline \end{tabular}
\end{table}

\newpage
\subsection*{``True'' formation energies}

The ``true'' formation energies (composition preserving \cite{Mishin}) are defined
from the ``raw'' formation energies (composition unpreserving \cite{Mishin}) as
\begin{equation}
\label{Eq:dE} \begin{array}{c}
    E_{\A}^{(i)} \equiv \mu_i -\mu_\A^\pure, \\\\
    E_{\A}^{(s)} \equiv \omu_s + \mu_\Mg^\pure - \mu_\A^\pure,
\end{array}
\end{equation}
where $\mu_\A^\pure$ and $\mu_\Mg^\pure$ are the chemical potentials
(energies per atom) of $A$- and Mg-atoms in pure $A$- and Mg-solids
at zero temperature, respectively.

\subsection*{Low-solubility formation energies}

In Ref. \onlinecite{SolubTi}, a quantity called ``low-solubility formation energy'', was shown to determine the dilute solubility
in binary alloys through the temperature exponential factor.
The quantity can be generalized to the case of Mg-B-$A$ alloy as (compare with Eq. (29) in Ref. \onlinecite{SolubTi}):
\begin{equation}  \label{Eq:dEGSIII}
  E_\sol^\alpha(A) \equiv \Delta E^\alpha_{\sc/\at}/x_\A^\sc
\end{equation}
with $\alpha=(i),(s)$ and
\begin{equation}
  \label{Eq:dEtern}
  \begin{array}{c}
    \Delta  E^{(i)}_{\sc/\at}=
    \frac{E_\sc[\A^{(i)}\Mg_{n_\ssc}\B_{m_\ssc}]}{1+n_\sc+m_\sc}-E_\at^\mix(x_\A^\sc,x_\B^\sc,x_\Mg^\sc),\\\\
    \Delta  E^{(s)}_{\sc/\at}=
    \frac{E_\sc[\A^{(s)}\Mg_{n_\ssc-1}\B_{m_\ssc}]}{n_\sc+m_\sc}-E_\at^\mix(x_\A^\sc,x_\B^\sc,x_\Mg^\sc),
  \end{array}
\end{equation}
where $E_\at^\mix(x_\A^\sc,x_\B^\sc,x_\Mg^\sc)$ is the energy of
a the three phase mixture at the same composition of the supercell
(see Table \ref{TAB:CscIII}):
\begin{equation}
  \label{Eq:Emix} E_\at^\mix(x_\A^\sc,x_\B^\sc,x_\Mg^\sc)=
  X_1\frac{E_\uc[\Mg_{n_\suc}\B_{m_\suc}]}{n_\suc+m_\suc}+ X_2E_\at^{(2)}+ X_3E_\at^{(3)}.
\end{equation}
The coefficients $X_k$ ($k$=1,2,3) are determined from the following linear
system of equations
\begin{equation}
  \label{Eq:X-III} \left\{
  \begin{array}{c}
    x_\A^\sc=\,\,\,\,\,\,\,\,\,\,\,\,\,\,\,\,\,\,\,\,\,\,\,\,\,\,\,\,\,\,\,\,\,\,
    X_2x_\A^{(2)}+X_3 x_\A^{(3)} \\
    x_\B^\sc=X_1 \frac{m_\suc}{n_\suc+m_\suc}+X_2 x_\B^{(2)}+X_3  x_\B^{(3)}\\
    x_\Mg^\sc=X_1 \frac{m_\suc}{n_\suc+m_\suc}+X_2 x_\Mg^{(2)}+X_3 x_\Mg^{(3)},
  \end{array}
  \right.
\end{equation}
where $E_\at^{(k)},x_\A^{(k)},x_\B^{(k)},x_\Mg^{(k)}$ ($k$=2,3) are the
energies (per atom) and stoichiometric compositions of the two other ground states
that, together with Mg$_{n_\suc}$B$_{m_\suc}$, form the {\it convex-hull triangle}
containing the point $(x_\A^\sc,x_\B^\sc,x_\Mg^\sc)$ in the ternary phase diagram.
For example, MgB$_{2}$, Mg, and LiB$_3$ are the three ground states
surrounding Mg$_7$B$_{16}$Li (a supercell of MgB$_2$ with a substitutional Li atom)
as shown in Fig. \ref{Fig:PhDi-31}.
The coefficient $X_k$ ($k$=1,2,3) represents the fraction of the $k$-th phase in the mixture.
The quantities $\Delta E^\alpha_{\sc/\at}$ defined in Eq. (\ref{Eq:dEtern})
are the supercell formation energies (per atom) determined with respect to the mixture of ground states \cite{SolubTi}.
$E_\sol^\alpha(A)$ and $\Delta E^\alpha_{\sc/\at}$ can not be negative.
If they do, it indicates that the list of ground states is incomplete, and a better phase diagram
should be established (the missed ground state can be supercell itself) \cite{SolubTi}.

In analogy of with binary alloys \cite{SolubTi},
the ternary alloy ``low-solubility formation energy'' $E_\sol^\alpha(A)$ is shown to determine the
solubility of $A$ in the low-solubility limit (Section \ref{SEC:solubility.theory}).

\section{Free energy of Mg-B-$A$ solid solution} \label{SEC:F1}

The Gibbs free energy per unit cell ({\it ``uc''}) of the Mg-B-$A$ solid solution
is determined within the mean-field approximation:
\begin{equation} \label{Eq:F1uc}
\begin{array}{c}
  G_{\uc}^{(1)}[\{c_i,c_s\},T]=E_{\uc}[\Mg_{n_\suc}\B_{m_\suc}]+
  \Delta g_\sI+\Delta g_\sS, \\\\
  \Delta g_\sI=\sum_{i} \nu_{i}\{c_i\mu_i+k_B T[c_i\ln c_i+(1-c_i)\ln (1-c_i)]\}, \\\\
  \Delta g_\sS=\sum_{s} \nu_{s}\{c_s\omu_s+k_BT[c_s\ln c_s+(1-c_s)\ln (1-c_s)]\},
\end{array}
\end{equation}
where $k_\B$ is a Boltzmann constant, $T$ is the temperature, and
$E_{\uc}[\Mg_{n_\suc}\B_{m_\suc}]$ represents the energy (enthalpy) of initial
$\Mg_{n_\suc}\B_{m_\suc}$ unit cell without $A$-solutes; the summations over $i$
and $s$ are over all the inequivalent types of
interstitial and substitutional positions in unit cell; $c_i$ and
$c_s$ are equal to the \emph{site}-concentrations of $A$-atoms at
each interstitial ($i$-) and substitutional ($s$-) type, respectively.
Thus, we assume that the concentrations in the equivalent positions are equal in the disordered state.
The ``raw'' formation energies $\mu_i$ and $\omu_s$ are
introduced in Sec. \ref{SEC:FormEn}.
Although, the mean-field approximation neglects correlations,
it should work well when the deviation from stoichiometry is small (see Sec. 19 in Ref. \onlinecite{KrivSm64}).
In addition, we neglect solute-solute interactions that might be important especially for high solute concentrations.
In conclusion, our model is similar to Wagner-Schottky model of a system of
non-interacting particles \cite{WagnerSchottky30}.

For given concentrations $\{c_i,c_s\}$,
the total number of atoms per unit cell, $N^\at_\uc$,
the total concentrations of $A$-interstitials, $x_{\A^{(i)}}^{(1)}$,
$A$-substitutional, $x_{\A^{(s)}}^{(1)}$,
and the total concentration of $A$-atoms, $x_\A^{(1)}$,
are determined as
\begin{equation}\label{Eq:CaI1}
\begin{array}{c}
    N^\at_\uc=\sum_{i}\nu_{i}c_i+n_\uc+m_\uc,\\\\
    x_{\A^{(i)}}^{(1)}= \sum_{i} \nu_{i}c_i / N^\at_\uc ,\texttt{ }
    x_{\A^{(s)}}^{(1)}=\sum_{s}\nu_{s}c_s / N^\at_\uc , \\\\
    x_\A^{(1)}=x_{\A^{(i)}}^{(1)}+x_{\A^{(s)}}^{(1)}.
\end{array}
\end{equation}

At given temperature and concentration $x_\A^{(1)}$,
the Gibbs free energy per atom $G_\at^{(1)}[x_\A^{(1)},T]$ is
determined by minimizating Eq. (\ref{Eq:F1uc}) with respect to
$\{c_i\}$ and $\{c_s\}$:
\begin{equation}\label{Eq:Min2}
  G_\at^{(1)}[x_\A^{(1)},T]= \min_{\{c_i,c_s\}} \left[ G_{\at}^{(1)}[\{c_i,c_s\},T] \right] _{x_\A^{(1)}},
\end{equation}
where
\begin{equation}\label{Eq:F1at}
  G_{\at}^{(1)}[\{c_i,c_s\},T]=G_{\uc}^{(1)}[\{c_i,c_s\},T]/N^\at_\uc.
\end{equation}
The solution $\{c_i,c_s\}$ defines the equilibrium distribution
of interstitial and substitutional $A$-solutes in the Mg$_{n_\uc}$B$_{m_\uc}$ solvent.

In the case of small concentrations of interstitials, the minimization can be
done with the Lagrange multiplier method, obtaining:
\begin{equation}
  \begin{array}{c}
    c_i=\left[ 1+\exp \frac{\mu_i+\mu(1-x_\A^{(1)})}{k_BT}
      \right]^{-1}, \texttt{ } c_s=\left[ 1+\exp \frac{\mu_s+\mu}{k_BT}
      \right]^{-1}.
  \end{array}
\end{equation}
where the Lagrange multiplier $\mu$ is determined from the following equation
(derived from Eq. (\ref{Eq:CaI1})):
\begin{equation}
  x_\A^{(1)}(n_{\uc}+m_{\uc})=\sum_{i}
  \nu_{i}c_i(1-x_\A^{(1)})+\sum_{s} \nu_{s}c_s.
\end{equation}

\section{Solubility}
\label{SEC:solubility.theory}

According to Nernst's theorem, either a single compound or a phase separation of compounds at correct stoichiometry
can be present at equilibrium at zero temperature.
At finite temperatures, the composition of phases can differ from stoichiometry
through solution because of the entropic promotion (Sec. \ref{SEC:solubility.results} and Ref. \onlinecite{SolubTi}).
At a given temperature, the solubility of $A$-atoms in a compound
is defined as the maximum homogeneously achievable concentration of $A$-atoms,
without the formation of a new phase.

To calculate the solubility \cite{note2}, we consider the
Gibbs free energy $G_\at^\mix$ of the mixture of three phases with a given
general ({\it ``gen''}) composition
$x^\gen\equiv(x^\gen_\A,x^\gen_\B,x^\gen_\Mg)$.
The first phase is the substitutional and/or interstitial solid
solution with atomic concentration $x_\A^{(1)}$ of element $A$ in
Mg$_{n_\uc}$B$_{m_\uc}$.
At zero temperature, the two other phases and Mg$_{n_\uc}$B$_{m_\uc}$
form a triangle containing the point
$x^\gen$ in the ternary phase diagram.
At a finite temperature, the  free energy $G_\at^\mix$ is the generalization of Eq. (\ref{Eq:Emix}):
\begin{equation}\label{Eq:Fmix}
\begin{array}{c}
    G_\at^\mix[x_\A^{(1)},T] \simeq
    X_1[x_\A^{(1)},T]G_\at^{(1)}[x_\A^{(1)},T] + \\\\
    X_2[x_\A^{(1)},T]E_\at^{(2)}+ X_3[x_\A^{(1)},T]E_\at^{(3)},
\end{array}
\end{equation}
where $G_\at^{(1)}[x_\A^{(1)},T]$ is given by Eq. (\ref{Eq:Min2}),
and the second and third phases are assumed to be stoichiometric, so that
their free energies can be approximated by their ground state energies (enthalpies)
$E_\at^{(2)}$ and $E_\at^{(3)}$, respectively.
The approximation does not affect much the results, as we are interested in the small solubility regime of $A$.
Similarly to Eq. (\ref{Eq:X-III}),
the fractions $X_k$ are determined by solving the system:
\begin{equation} \label{Eq:X-III-2} \left\{
  \begin{array}{c}
    x^\gen_\A=X_1 x_\A^{(1)}+X_2
    x_\A^{(2)}+X_3
    x_\A^{(3)} \\
    x^\gen_\B=X_1 x_\B^{(1)}+X_2
    x_\B^{(2)}+X_3
    x_\B^{(3)}\\
    x^\gen_\Mg=X_1 x_\Mg^{(1)}+X_2
    x_\Mg^{(2)}+X_3 x_\Mg^{(3)},
\end{array}
\right.
\end{equation}
where the second ``(2)'' and third ``(3)'' phases are at stoichiometry.
In addition, we have
\begin{equation}\label{Eq:C1}
    \begin{array}{c}
        x_\B^{(1)}=\frac{n_{\uc}}{n_{\uc}+m_{\uc}}\left(1-x_{\A^{(i)}}^{(1)}[x_\A^{(1)},T] \right), \\
        x_\Mg^{(1)}=1-x_\A^{(1)}-x_\B^{(1)},
    \end{array}
\end{equation}
where $x_{\A^{(i)}}^{(1)}[x_\A^{(1)},T]$ is the equilibrium concentration
of $A$-interstitials in the first phase given by Eqs. (\ref{Eq:CaI1}-\ref{Eq:Min2}) at
for chosen total concentration $x_\A^{(1)}$.
The minimization of $G_\at^\mix$ with respect to
$x_\A^{(1)}$ gives the solubility $x_\A^{(1)}(T)$ in
the first phase ``(1)'':
\begin{equation}\label{Eq:FmixEQ}
    G_\at^\mix[x_\A^{(1)}(T),T]=\min_{x_\A^{(1)}}
    G_\at^\mix[x_\A^{(1)},T].
\end{equation}
This procedure, equivalent to the common-tangent method,
is the generalization of the approach developed in Ref. \onlinecite{SolubTi} to the case of ternary alloys.
The method is
somehow different from that developed in Ref. \onlinecite{SigliPROC}.
Although, the regular solution model used in Ref. \onlinecite{SigliPROC}
and the presented ideal solution model coincide in the low solubility regime,
the consideration of only two ground states ({\it ``gs''}) by the authors of Ref. \onlinecite{SigliPROC},
differs from our approach requiring the knowledge of the whole ternary stability.
This is because the disordered Mg-B-$A$ solution does not generally
belong to the Mg-B$\leftrightarrow$$gs$$^{(2)}$ or
Mg-B$\leftrightarrow gs^{(3)}$ lines in the ternary phase diagram.
Thus, we have to consider the phase mixture of Mg$_{n_\uc}$B$_{m_\uc}$,
$gs^{(2)}$, and , $gs^{(3)}$ to guarantee accurate estimation of the solubility.

\subsection*{Low-solubility approximation}

In order to get the analytical expression for equilibrium solubilities from Eq. (\ref{Eq:FmixEQ}),
further approximations are required:
(a) the equilibrium concentration of solute $A$ is small and
(b) only substitutional or interstitial positions of one type $\alpha$ are occupied
(this approximation will be {\bf eliminated} at the end of section).
Thus, from Eqs. (\ref{Eq:Fmix},\ref{Eq:FmixEQ})) we obtain:
\begin{equation} \label{Eq:LS1}
    \frac{\partial G_\at^\mix}{ \partial x_\A^{(1)}} \approx
    \frac{\partial E_\at^\mix}{ \partial x_\A^{(1)}} +
    X_1\frac{\partial G_\at^{(1)}[x_\A^{(1)},T]}{ \partial
    x_\A^{(1)}}=0,
\end{equation}
where
\begin{equation}\label{Eq:LS15}
    E_\at^\mix[x_\A^{(1)},x^\gen]= X_1E_\at^{(1)}+
    X_2E_\at^{(2)}+ X_3E_\at^{(3)},
\end{equation}
and the fractions $X_k=X_k[x_\A^{(1)},x^\gen]$ come from Eqs. (\ref{Eq:X-III-2},\ref{Eq:C1}).
Note that the {\it circa} ($\approx$) in Eq. (\ref{Eq:LS1})
corresponds to approximation (a) applied to Eqs.
(\ref{Eq:F1uc},\ref{Eq:F1at}):
\begin{equation}
    \frac{\partial X_1}{\partial x_\A^{(1)}}G_\at^{(1)} \approx
    \frac{\partial X_1}{\partial x_\A^{(1)}}E_\at^{(1)}.
\end{equation}
Equations (\ref{Eq:X-III-2},\ref{Eq:C1}) lead to:
\begin{equation}\label{Eq:LS25}
    \frac{\partial X_k}{ \partial x_\A^{(1)}} \approx
    -X_1 \frac{\partial X_k}{ \partial x^\gen_\A}
\end{equation}
and through Eq. (\ref{Eq:LS15}):
\begin{equation}\label{Eq:LS2}
    \frac{\partial E_\at^\mix}{ \partial x_\A^{(1)}} \approx
    -X_1 \frac{\partial E_\at^\mix}{ \partial
    x^\gen_\A}.
\end{equation}
By using the expression of $G^{(1)}$ from Eq. (\ref{Eq:F1uc}), we have
\begin{equation}\label{Eq:LS3}
\frac{\partial G_\at^{(1)}}{ \partial x_\A^{(1)}} \approx \mu_\alpha
+ k_BT \ln \frac{c_\alpha}{1-c_\alpha},
\end{equation}
where the {\it circa} ($\approx$) corresponds to approximation (a) \cite{note3}.

Equation (\ref{Eq:LS1}) can be solved with respect to $c_\alpha$
with the help of Eqs. (\ref{Eq:LS2},\ref{Eq:LS3}) as
\begin{eqnarray}
  c_\alpha&=&\left [1 + \exp \frac{E_\sol^\alpha(A)}{k_BT} \right ]^{-1} \nonumber \\
  &\simeq& \left. \exp \left(-\frac{E_\sol^\alpha(A)}{k_\B T} \right )  \right|_{k_\B T \ll E_\sol^\alpha(A)}, \label{Eq:LS4}
\end{eqnarray}
where
\begin{eqnarray}
 && E_\sol^\alpha(A) = \mu_\alpha-\frac{\partial E_\at^\mix}{ \partial x^\gen_\A} \nonumber \\
  &&=\frac{1}{x_\A^\gen} \left[ E_\at^{(1)} + \mu_\alpha x_\A^\gen - \left(E_\at^{(1)} + \frac{\partial E_\at^\mix}{ \partial x^\gen_\A} x_\A^\gen \right) \right ] \nonumber \\
  && = \frac{1}{x_\A^\gen} \left[ E^{(1)}_\at(x^\gen) - E_\at^\mix(x^\gen) \right ].\label{Eq:LS5}
\end{eqnarray}
Both Equations (\ref{Eq:LS5}) and (\ref{Eq:dEGSIII}) define the same quantity:
the low-solubility formation energy, $E_\sol^\alpha(A)$,  determining the low-solubility of
$A$-solutes in $\alpha$ type positions.
In addition, by using  Eqs. (\ref{Eq:CaI1})-(\ref{Eq:LS4})\cite{note3}
the total equilibrium concentration of solute $A$ in the phase (1) becomes
\begin{equation}\label{Eq:LS7}
    x_\A^{(1)} \approx \frac{\nu c_\alpha}{n_{\uc}+m_{\uc}},
\end{equation}
where $c_\alpha$ is determined by Eq. (\ref{Eq:LS4}).

Approximation (b) can be relaxed if the various types of
substitutional and interstitial positions are occupied independently.
This is expected to be true in the low solute concentration limit.
Thus, expression (\ref{Eq:LS7}) can be integrated out through the various types of positions, leading to
\vspace{-4mm}
\begin{equation}\label{Eq:LS9}
  x_\A^{(1)} \approx \frac{\sum_{i} \nu_{i}c_i+\sum_{s}
    \nu_{s}c_s}{n_{\uc}+m_{\uc}},
\end{equation}
where $c_i$ and $c_s$ are obtained from a set of Eq. (\ref{Eq:LS4}),
by using the corresponding $E_\sol^{(i)}(A)$ and $E_\sol^{(s)}(A)$ for each type of defects.

\section{First-principles calculations}
\label{SEC:1pr}

The first-principles calculations of energies are
performed by
using our high-throughput quantum calculations framework {\small AFLOW}
\cite{SC13,SC20,MS1,Kolmogorov_Lithium_Borides}
and the software {\small VASP} \cite{kresse1993}.
We use projector
augmented waves (PAW) pseudopotentials \cite{bloechl994} and exchange-correlation functionals
as parameterized by Perdew and Wang \cite{PW91} for the generalized
gradient approximation (GGA). Simulations are carried out without
spin polarization (not required for the elements under
investigation), at zero temperature, and without zero-point motion.
All structures are fully relaxed (shape and volume of the cell and
internal positions of the atoms). The effect of lattice vibrations
is omitted. Numerical convergence to within about 1 meV/atom is
ensured by enforcing a high energy cut-off (414 eV) and dense 4,500
{\bf k}-point meshes.

\subsection*{Ground states determination}
\label{SEC:GS}

The calculation of solubility of $A$ in Mg-B compounds
requires the knowledge of the relevant ground states in the ternary Mg-B-$A$ system.
The  systems under investigations have not been
well characterized experimentally or theoretically, and
only the binary Mg-B, B-$A$, and Mg-$A$ systems have been studied.
Hence, we performed additional high-throughput searches to determine if further
ground states exist in the three binary systems \cite{SC13,SC20,MS1,Kolmogorov_Lithium_Borides}.
Based on the knowledge of the binary systems,
we built \cite{qhull} the ternary ground state phase
diagrams for Mg-B-$A$, with the expectation that no
{\it missed} ternary ground state is relevant to the solubility of $A$ (Ref. \onlinecite{note4}).

\vspace{-4mm}
\begin{table}[h]
\caption{\small
  The formation energies $\Delta E^\bin_\at$ (Eq.
  (\ref{Eq:dEbin})) for Mg-B, B-$A$, and Mg-$A$ binary ground states
  ($A$=Li, Be, Na, Mg, K, Ca, Rb, Sr, Cs, Ba).
  The symbol (+) indicates possible ground states never observed experimentally (CaB$_{4}$ and RbB$_{4}$).
  The symbol (-) indicates the experimentally observed
  Sr$_{9}$Mg$_{38}$, which was not confirmed to be a ground states by
  first-principle calculations.} \label{Tab:dEbin} \footnotesize
\begin{tabular}{c|c|c|c|c|c} \hline\hline
  &   $\Delta E^\bin_\at$    &   Proto-  &   Space   &   Pearson   &   Ref.    \\
  &   (eV/at.)    &   type\cite{Pauling}  &   group\cite{IntTabs} &       &       \\  \hline
  Li$_3$B$_{14}$  &   -0.219  &   Li$_3$B$_{14}$  &   I$\overline{4}$2d(122)  &   tI160   &   \onlinecite{Li3B14} \\
  LiB$_3$ &   -0.235  &   LiB$_3$ &   P4/mbm(127) &   tP20    &   \onlinecite{LiB3}   \\
  Li$_8$B$_7$ &   -0.216  &   Li$_8$B$_7$ &    P$\overline{6}$m2(187)   &   hP15 &   \onlinecite{Kolmogorov_Lithium_Borides}  \\
  Be$_3$B$_{50}$  &   -0.032  &   Be$_3$B$_{50}$  &   P4$_2$/nnm(134) &   tP53    &   \onlinecite{Be3B50} \\
  Be$_{1.11}$B$_3$    &   -0.096  &   Be$_{1.11}$B$_3$    &   P6/mmm(191) &   hP111   &   \onlinecite{BeB3}   \\
  MgBe$_{13}$ &   -0.009  &   NaZn$_{13}$ &   Fm$\overline{3}$c(226)  &   cF112   &   \onlinecite{MgBe13} \\
  NaB$_{15}$  &   -0.059  &   NaB$_{15}$  &   Imma(74)    &   oI64    &   \onlinecite{NaB15}  \\
  Na$_3$B$_{20}$  &   -0.070  &   Na$_3$B$_{20}$  &   Cmmm(65)    &   oC46    &   \onlinecite{Na3B20} \\
  MgB$_7$ &   -0.138  &   MgB$_7$ &   Imma(74)    &   oI64    &   \onlinecite{MgB7}   \\
  MgB$_{4}$   &   -0.152  &   MgB$_{4}$   &   Pnma(62)    &   oP20    &   \onlinecite{MgB4}   \\
  MgB$_{2}$   &   -0.151  &   AlB$_{2}$   &   P6/mmm(191) &   hP3 &   \onlinecite{MgB2}   \\
  KB$_{6}$    &   -0.043  &   CaB$_{6}$   &   Pm$\overline{3}$m(221)  &   cP7 &   \onlinecite{KB6}    \\
  CaB$_{6}$   &   -0.423  &   CaB$_{6}$   &   Pm$\overline{3}$m(221)  &   cP7 &   \onlinecite{CaB6}   \\
  CaB$_{4}$(+)&   -0.410  &   UB$_{4}$    &   P4/mbm(127) &   tP20    &   \onlinecite{CaB4}   \\
  CaMg$_{2}$  &   -0.128  &   MgZn$_{2}$  &   P6$_3$/mmc(194) &   hP12    &   \onlinecite{CaMg2}  \\
  RbB$_{4}$(+)   &   -0.163  &   UB$_{4}$    &   P4/mbm(127) &   tP20    &   - \\
  SrB$_{6}$   &   -0.464  &   CaB$_{6}$   &   Pm$\overline{3}$m(221)  &   cP7 &   \onlinecite{SrB6BaB6}   \\
  Sr$_{2}$Mg$_{17}$   &   -0.055  &   Th$_{2}$Ni$_{17}$   &   P6$_3$/mmc(194) &   hP38    &   \onlinecite{Sr2Mg17Ba2Mg17} \\
  Sr$_{9}$Mg$_{38}$(-)   &   -0.070  &   Sr$_{9}$Mg$_{38}$   &   P6$_3$/mmc(194) &   hP94    &   \onlinecite{Sr9Mg38}    \\
  Sr$_{6}$Mg$_{23}$   &   -0.085  &   Th$_{6}$Mg$_{23}$   &   Fm$\overline{3}$m(225)  &   cF116   &   \onlinecite{Sr6Mg23Ba6Mg23} \\
  SrMg$_{2}$  &   -0.114  &   MgZn$_{2}$  &   P6$_3$/mmc(194) &   hP12    &   \onlinecite{SrMg2BaMg2} \\
  BaB$_{6}$   &   -0.421  &   CaB$_{6}$   &   Pm$\overline{3}$m(221)  &   cP7 &   \onlinecite{SrB6BaB6}   \\
  Ba$_{2}$Mg$_{17}$   &   -0.072  &   Zn$_{17}$Th$_{2}$   &   R$\overline{3}$mh(166)  &   hR57    &   \onlinecite{Sr2Mg17Ba2Mg17} \\
  Ba$_{6}$Mg$_{23}$   &   -0.085  &   Th$_{6}$Mg$_{23}$   &   Fm$\overline{3}$m(225)  &   cF116   &   \onlinecite{Sr6Mg23Ba6Mg23} \\
  BaMg$_{2}$  &   -0.097  &   MgZn$_{2}$  &   P6$_3$/mmc(194) &   hP12    &   \onlinecite{SrMg2BaMg2} \\
  \hline\hline \end{tabular} \end{table}

\begin{figure}[h!]
  \includegraphics[width=0.51\textwidth]{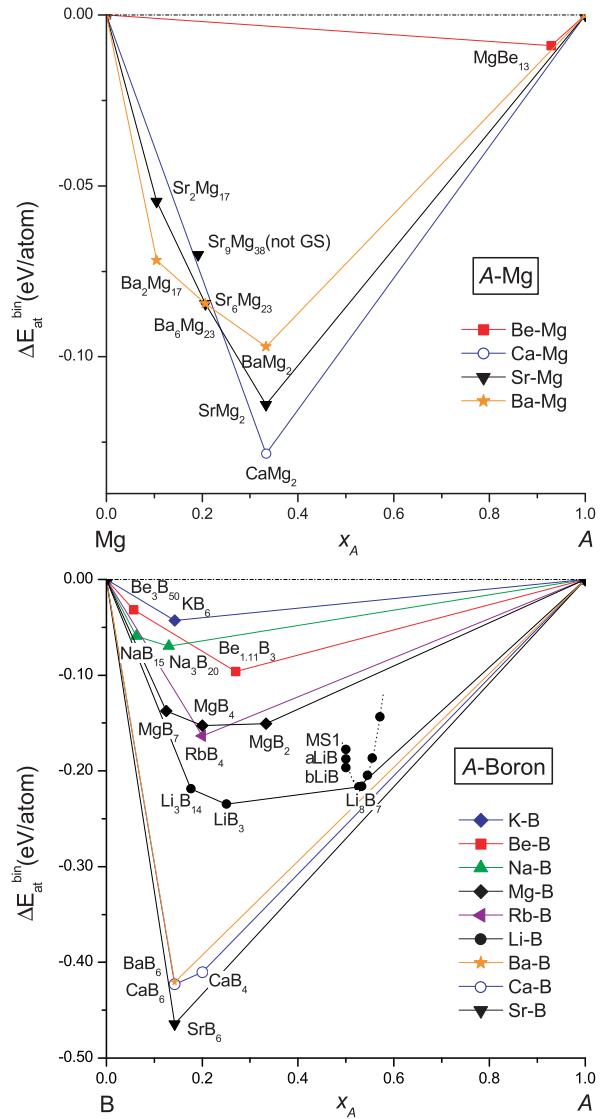}
  \vspace{-13mm}
  \caption{\small
    (Color online)
    Formation energies $\Delta E^\bin_\at$ for Mg-B, B-$A$, and Mg-$A$ binary ground states
    ($A$=Li, Be, Na, Mg, K, Ca, Rb, Sr, Cs, Ba) from Table \ref{Tab:dEbin}.
    The dashed parabola for the Li-B system indicates the linear chains phases of Ref. \onlinecite{Kolmogorov_Lithium_Borides}}
  \label{Fig:PhDi-2}
\end{figure}

The existence of the ground states for a given binary $A$-$B$ system is based on binary
{\it bulk} formation energy, which, for each phase $\phi$ with stoichiometry $A_{x_\ssA}B_{x_\ssB}$,
is determined with respect to pure $A$ and $B$ energies $E_\at(A)$ and $E_\at(B)$ as
\begin{equation}  \label{Eq:dEbin}
  \Delta E^\bin_\at(\phi) \equiv
  E_\at(\phi)-x_\sA E_\at(A)-x_\sB E_\at(B).
\end{equation}
The results are presented at Table \ref{Tab:dEbin} and in Figure
\ref{Fig:PhDi-2}. For each element $A$, the reference energy is
chosen to be the lowest among the pure fcc, bcc and hcp
energies\cite{SC15}. The reference energy for boron is taken to be
$\alpha$-boron (Refs. \onlinecite{Pauling,aLiB,bLiB,boron,MS1}).

All experimentally observed phases are confirmed except for Sr$_9$Mg$_{38}$ (P6$_3$/mmc).
We also find two possible new phases,
CaB$_{4}$ and RbB$_{4}$, both P4/mbm (\#127) and with UB$_{4}$ prototype
(CaB$_{4}$ was previously identified in {\it ab initio} online database\cite{CaB4}).
The formation energies for Ba$_{2}$Mg$_{17}$, Ba$_{6}$Mg$_{23}$,
BaMg$_{2}$, and CaMg$_{2}$ are similar to those of Ref. \onlinecite{ZKL07}
(reported as -0.079 eV, -0.089 eV, -0.088 eV, and -0.126 eV, respectively).
The small differences can be explained on the basis of different GGA pseudopotentials (PW91 versus PBE)
and different energy cut-offs (414 eV versus 360 eV).
Our results can not be compared with the thermodynamic discussion based on the semi-empirical Miedema method
reported in Ref. \onlinecite{WuGao05}, because we include further ground state prototypes other than those
used in the heat of formations fitting in Ref. \onlinecite{WuGao05}.

\begin{figure}[t!]
  \vspace{-18mm}
  \includegraphics[width=0.51\textwidth]{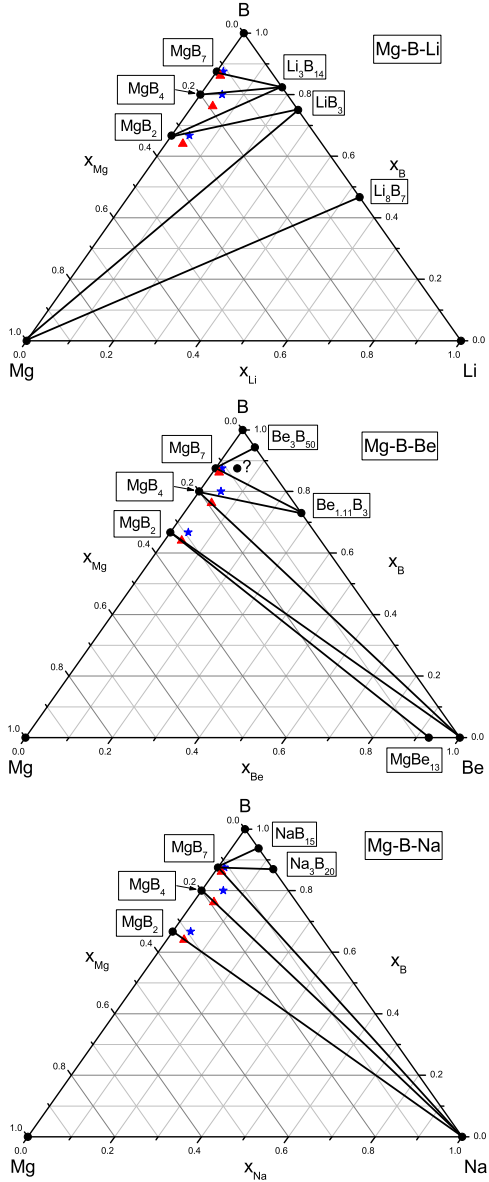}
  \vspace{-18mm}
  \caption{\small
    (color online) The ternary ground state phase diagrams of
    Mg-B-$A$ alloys ($A$=Li, Be, Na). The filled circles represent known
    ground states. The blue stars and red triangles represent the
    substitutional (Mg$_{7}$B$_{56}A$, Mg$_{3}$B$_{16}A$, Mg$_{7}$B$_{16}A$) and interstitial
    (Mg$_{8}$B$_{56}A$, Mg$_{4}$B$_{16}A$,
    Mg$_{8}$B$_{16}A$) supercells, respectively.
    Each supercell has been constructed from the most appropriate Mg-B compound
    (MgB$_2$, MgB$_4$, or MgB$_7$). The question mark in the Mg-B-Be
    system indicates the existence of new possible ternary ground state(s) (Sec. \ref{SEC:FormEn.results}).}
  \label{Fig:PhDi-31}
\end{figure}

\begin{figure}[thb]
  \vspace{-18mm}
  \includegraphics[width=0.51\textwidth]{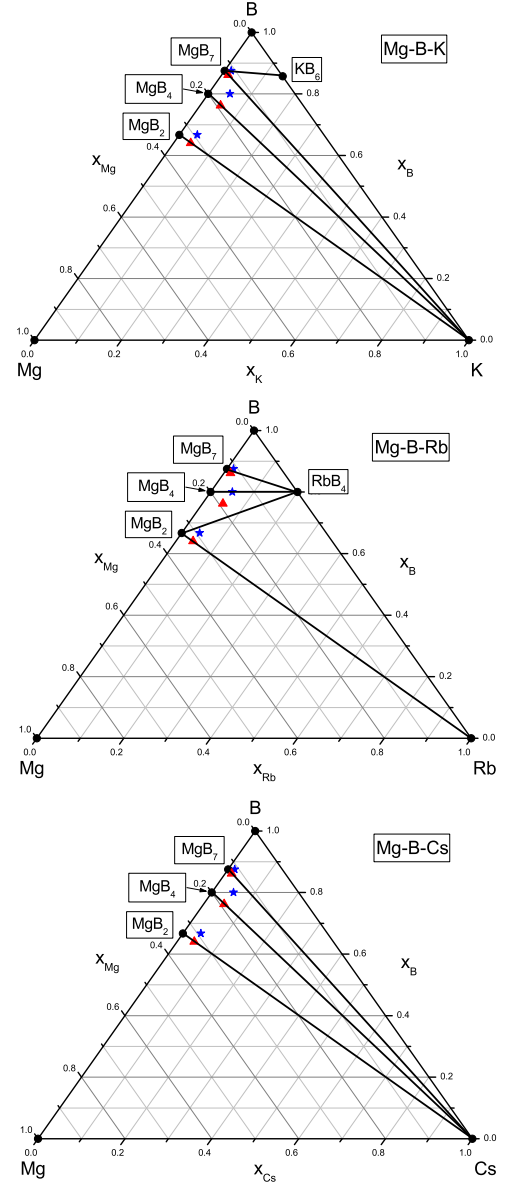}
 \vspace{-18mm}
  \caption{\small
    (color online) The ternary ground state phase diagrams of
    Mg-B-$A$ alloys ($A$=K, Rb, Cs). } \label{Fig:PhDi-32}
\end{figure}

\begin{figure}[thb]
   \vspace{-18mm}
 \includegraphics[width=0.51\textwidth]{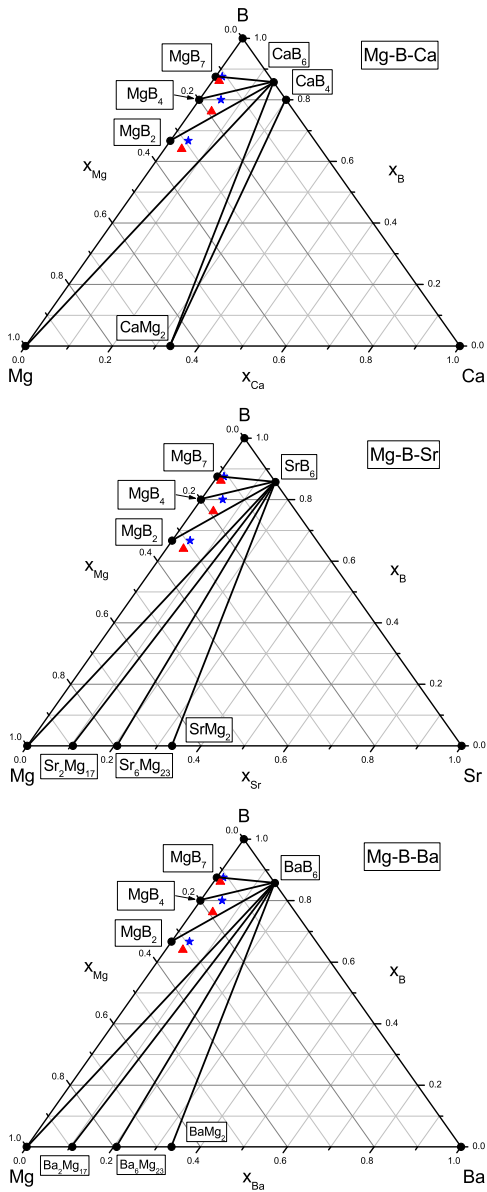}
 \vspace{-18mm}
  \caption{\small
       (color online) The ternary ground state phase diagrams of
    Mg-B-$A$ alloys ($A$=Ca, Sr, Ba). } \label{Fig:PhDi-33}
\end{figure}

%
%
%
%
%

The calculated ternary ground state phase diagrams for Mg-B-$A$
alloys are depicted in Figs. \ref{Fig:PhDi-31},\ref{Fig:PhDi-32},\ref{Fig:PhDi-33}.
Note that in each phase diagram, the red triangles and blue stars represent the
supercells with one interstitial or substitutional $A$ atom,
respectively. In the interstitial case, the supercells belong to
the lines Mg$_n$B$_m$$\leftrightarrow{\hspace{-1.5mm}A}$, while in the
substitutional case, the supercells belong to the lines parallel to
Mg$\leftrightarrow{\hspace{-1.5mm}A}$ and intersecting the Mg$_n$B$_m$
(constant B concentration).

\vspace{-5mm}
\subsection*{Formation energies numerical results}
\label{SEC:FormEn.results}

\begin{table}
  \caption{\small
    Interstitial (i) and substitutional (s) quantities for alkali and alkaline earth solute
    elements $A$ in MgB$_2$, MgB$_4$, and MgB$_7$ compounds.:
    ``raw'' formation energies $\mu_i$, $\omu_s$ (Eq. (\ref{Eq:mIII})),
    ``true'' formation energies $E_{\A}^{(i)}$, $E_{\A}^{(s)}$ (Eq. (\ref{Eq:dE})) and
    low-solubility formation energies $E_\sol^{(i)}$, $E_\sol^{(s)}$ (Eqs. (\ref{Eq:dEGSIII})).
    In each case, the presented value is the lowest one among all the possible interstitial or substitutional positions.} \label{Tab:dE} \small
  \begin{tabular}{c|c|c|c|c|c|c|c} \hline\hline
    & &  \multicolumn{2}{c|}{MgB$_2$}   & \multicolumn{2}{c|}{MgB$_4$}   & \multicolumn{2}{c}{MgB$_7$} \\
    \hline
    $A$   & (eV) & (i) & (s)  & (i) & (s) & (i) & (s) \\
    \hline
    Li  &   $\mu_i,\omu_s$   &   0.542   &   -0.111  &   0.234   &   -0.109  &   -0.959  &   -0.055  \\
    &   $E_{\A}^{(i,s)}$  &   2.444   &   0.310   &   2.136   &   0.312   &   0.943   &   0.366   \\
    &   $E_\sol^{(i,s)}$   &   2.716   &   0.574   &   2.654   &   0.686   &   1.657   &   0.506   \\\hline
    Be  &   $\mu_i,\omu_s$   &   -2.665  &   -0.876  &   -0.506  &   -0.710  &   -1.577  &   -2.100  \\
    &   $E_{\A}^{(i,s)}$ &   1.042   &   1.350   &   3.201   &   1.516   &   2.130   &   0.126   \\
    &   $E_\sol^{(i,s)}$   &   1.054   &   1.21    &   3.201   &   1.256   &   2.181   &   -0.557  \\\hline
    Na  &   $\mu_i,\omu_s$   &   5.623   &   1.839   &   2.409   &   1.285   &   1.999   &   0.869   \\
    &   $E_{\A}^{(i,s)}$  &   6.932   &   1.666   &   3.718   &   1.113   &   3.308   &   0.697   \\
    &   $E_\sol^{(i,s)}$   &   6.944   &   1.526   &   3.718   &   0.802   &   3.308   &   0.147   \\\hline
    K   &   $\mu_i,\omu_s$   &   7.022   &   4.674   &   2.984   &   3.001   &   3.453   &   2.077   \\
    &   $E_{\A}^{(i,s)}$  &   8.061   &   4.232   &   4.023   &   2.559   &   4.492   &   1.635   \\
    &   $E_\sol^{(i,s)}$   &   8.073   &   4.092   &   4.023   &   2.248   &   4.492   &   0.838   \\\hline
    Ca  &   $\mu_i,\omu_s$   &   2.833   &   -0.295  &   1.042   &   -1.107  &   1.251   &   -1.941  \\
    &   $E_{\A}^{(i,s)}$  &   4.751   &   -0.732  &   2.960   &   -0.671  &   3.169   &   -1.504  \\
    &   $E_\sol^{(i,s)}$   &   6.364   &   1.747   &   4.993   &   1.218   &   5.453   &   0.358   \\\hline
    Rb  &   $\mu_i,\omu_s$   &   9.463   &   6.572   &   3.733   &   3.957   &   4.751   &   3.312   \\
    &   $E_{\A}^{(i,s)}$  &   10.393  &   6.021   &   4.663   &   3.406   &   5.681   &   2.761   \\
    &   $E_\sol^{(i,s)}$   &   10.406  &   5.981   &   4.862   &   3.460    &   6.047   &   2.477   \\\hline
    Sr  &   $\mu_i,\omu_s$   &   5.514   &   1.761   &   1.334   &   0.221   &   2.149   &   -0.960  \\
    &   $E_{\A}^{(i,s)}$  &   7.138   &   1.618   &   2.958   &   0.364   &   3.773   &   -0.817  \\
    &   $E_\sol^{(i,s)}$   &   9.042   &   3.802   &   5.282   &   2.544   &   6.348   &   1.331   \\\hline
    Cs  &   $\mu_i,\omu_s$   &   4.663   &   8.436   &   4.559   &   5.064   &   6.488   &   5.131   \\
    &   $E_{\A}^{(i,s)}$  &   5.523   &   7.815   &   5.419   &   4.444   &   7.349   &   4.510   \\
    &   $E_\sol^{(i,s)}$   &   5.536   &   7.675   &   5.419   &   4.134   &   7.349   &   3.411   \\\hline
    Ba  &   $\mu_i,\omu_s$   &   2.450   &   3.959   &   1.365   &   1.084   &   3.118   &   0.335   \\
    &   $E_{\A}^{(i,s)}$  &   4.373   &   3.518   &   3.288   &   1.525   &   5.041   &   0.777   \\
    &   $E_\sol^{(i,s)}$   &   5.971   &   5.99    &   5.306   &   3.400 &   7.309   &   2.624   \\
    \hline\hline \end{tabular} \end{table}

\begin{figure}[h!]
  \includegraphics[width=0.51\textwidth]{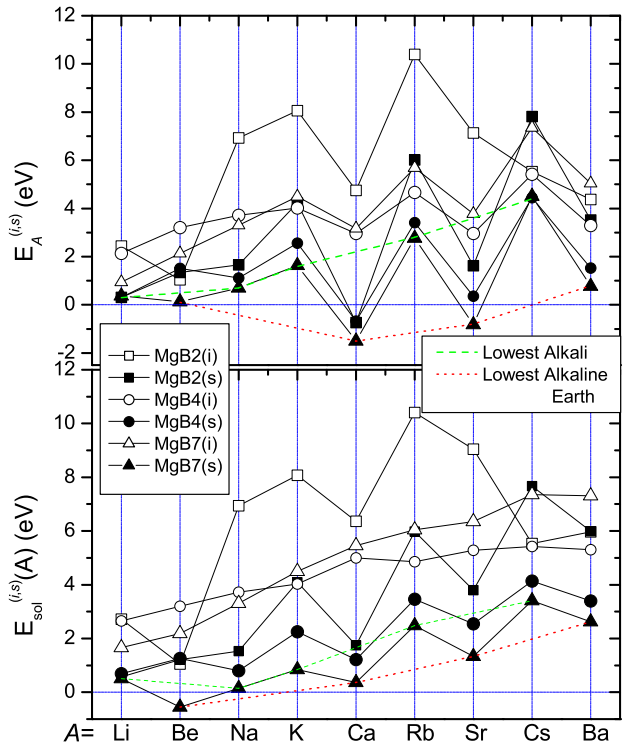}
  \vspace{-10mm}
  \caption{\small
    (Color online) Representation of the formations energies from Table \ref{Tab:dE}.
    Dashed (green) and dotted (red) lines connect the lowest formation energies for alkali and alkaline earth
    elements, respectively. } \label{Fig:dE}
\end{figure}

The calculated parameters of the model
(``raw'' $\mu_i$, $\omu_s$ (Eq. (\ref{Eq:mIII})), ``true''- $E_{\A}^{(i)}$, $E_{\A}^{(s)}$ (Eq. (\ref{Eq:dE}))
and low-solubility $E_\sol^{(i)}$, $E_\sol^{(s)}$ (Eqs. (\ref{Eq:dEGSIII})) formation energies) are presented in Table
\ref{Tab:dE} and in Fig. \ref{Fig:dE}.

The values of the low-solubility energies reported in Table \ref{Tab:dE} suggest the following.
(a) Only $E_\sol^{(s)}$ for substitutional Be in MgB$_7$ was found
to be negative, indicating the existence of unknown ground state(s)
within the triangle
MgB$_7\leftrightarrow{\hspace{-1.5mm}}$Be$_3$B$_{50}\leftrightarrow{\hspace{-1.5mm}}$Be$_{1.11}$B$_3$.
This fact is summarized by the question mark in Fig.
\ref{Fig:PhDi-31}.
(b) Generally, the substitutional systems have lower formation
energy than the interstitial systems.
Exceptions are Be, Cs, and Ba in MgB$_2$;
(c) Although oscillating, the formation energies tends to increase with the element number, as shown in Fig. \ref{Fig:dE}.
Regular oscillations are observed for all substitutionals systams starting with Na.
The lower and higher boundaries of such  correspond to alkaline earth and alkali elements,
respectively.
(d) Na, Ca, and Li substitutionals in MgB$_7$ have the lowest
formation energies (0.147 eV, 0.358 eV, and 0.506 eV, respectively), resulting in
high solubility (Sec. \ref{SEC:solubility.results}).
(e) In general for substitutionals systems, the higher boron contents the
lower formation energy (except Li and Be).
(f) The corresponding ``true''
$E_\A^{(i,s)}$ and low-solubility $E_\sol^{(i,s)}(A)$ formation
energies generally demonstrate a similar behavior as functions of $A$.
Exceptions are $E_{\A}^{(s)}<0$ for $A$=Ca and Sr and $E_\sol^{(s)}(A)<0$ for $A$=Be.
The difference is due to the consideration of
all the known ground states for the determination of  $E_\sol^{(i,s)}$
and not only pure Mg, B and $A$ (as for $E_\A^{(i,s)}$).

For Li-Mg-B, in Ref. \onlinecite{BernMass06} the formation energies were
obtained from first principles for interstitial and substitutional
Li in boron-rich MgB$_2$ (3.00 eV and 0.27 eV, respectively). The
results of Ref. \onlinecite{BernMass06} are different form our
low-solubility formation energies $E_\sol^{(i)}$(Li)=2.72 eV and
$E_\sol^{(s)}$(Li)=0.57 eV (see Table \ref{Tab:dE}), because only
the hexagonal $\alpha$-LiB phase was considered as a ground state in Ref. \onlinecite{BernMass06}
and the other boron-rich phases were reported to be unstable
(the $\alpha$-LiB was refined in
Ref. \onlinecite{Kolmogorov_Lithium_Borides} by variational
minimization of the Li and B concentration as Li$_8$B$_7$).
However, in our high-throughput framework, we found LiB$_3$ and Li$_3$B$_{14}$ to be
stable \cite{LiBBB}, and the approriate analysis of the formation energies should
be done with respect to the triangle MgB$_2$$\leftrightarrow{\hspace{-1.5mm}}$LiB$_3$$\leftrightarrow{\hspace{-1.5mm}}$Mg as
shown in Fig. \ref{Fig:PhDi-31}.
In the low-solubility limit of Li, since the supercell concentration is close to MgB$_2$,
the lever rule of phase decomposition of the mixtures does not
cause significant errors if the energies are calculated with respect to other references,
and therefore the results of  Ref. \onlinecite{BernMass06} are somehow similar.

For Na-Mg-B, it can be shown that the substitutional
Na formation energy in magnesium-rich MgB$_2$ (1.74 eV) obtained in
Ref. \onlinecite{BernMass06} is equivalent to our ``true'' formation
energy for substitutional Na in MgB$_2$ ($E_\texttt{\footnotesize Na}^{(s)}=1.67$ eV, see Table \ref{Tab:dE}).
The observed small numerical difference
between two results can be attributed to the different energy
cutoffs for the basis set (414 eV versus 312 eV)
\cite{note5}.

\section{Solubility results}
\label{SEC:solubility.results}

Solubility results are presented in Table \ref{TAB:solub}
(the low-solubility approximation values are shown in brackets).
We report only values larger than 10$^{-6}$ (Li, Na, Ca, and K).
The values on Table \ref{TAB:solub} suggest the following.
(a) Na is the only element which has a substantial solubility:
Na in MgB$_7$ is $\sim$ 0.5-1 \% at $T=650-1000$ K.
(b) The substitutional solubility of Be in MgB$_7$ can not be
determined because the corresponding formation energy $E_\sol^{(s)}$
(see Eq. (\ref{Eq:dEGSIII})) is negative implying that the ground
state list is not complete. Thus the approach of Sec.
\ref{SEC:solubility.theory} is not applicable.
(c) In agreement with their lowest formation energies
$E_\sol^{(s)}$ (Table \ref{Tab:dE} and Fig. \ref{Fig:dE}),
substitutional Li, Na, and Ca
experience the largest solubilities among all studied systems;
(d) Due to their comsiderably high formation energies
$E_\sol^{(i)}$ (Table \ref{Tab:dE} and Fig. \ref{Fig:dE}),
all the investigated interstitial systems have negligible solubilities.
(e) The low-solubility approximation and the general theory agree within 5\%.

\begin{table} \small
  \caption{Calculated equilibrium interstitial $x_{\A^{(i)}}^{(1)}$ (i) and
    substitutional $x_{\A^{(s)}}^{(1)}$ (s) $A$-solute concentrations at
    temperatures $T=$300, 650, and 1000 K, in MgB$_2$, MgB$_4$, and
    MgB$_7$ compounds. The values obtained within the
    low-solubility approximation are reported in brackets.
    Zeros are used for values smaller than $10^{-6}$. For Be ``$\Delta E<0$''
    indicates that the corresponding formation energy
    $E_\sol^{(s)}$ (see Eq. (\ref{Eq:dEGSIII})) is negative implying
    that the ground state(s) list is not complete and
    the approach of Sec. \ref{SEC:solubility.theory} is not applicable.}
  \label{TAB:solub}
  \begin{tabular}{c|c|c|c|c|c|c|c} \hline\hline
    \multicolumn{2}{c|}{$x_{\A^{(i,s)}}^{(1)}$}     & \multicolumn{2}{c|}{MgB$_2$}   & \multicolumn{2}{c|}{MgB$_4$}   & \multicolumn{2}{c}{MgB$_7$} \\
    \hline
    $A$ & $T$ (K)   & i & s  & i & s & i & s \\
    \hline Li  &   300 &   0   &   0   &   0   &   0   &   0   &   0   \\
    &   650 &   0   &   1.3$\times 10^{-5}$ &   0   &   0   &   0   &   9.9$\times 10^{-6}$ \\
    &       &       &   (1.2$\times 10^{-5}$)   &       &       &       &   (9.9$\times 10^{-6}$)   \\
    & 1000  &   0   &   4.5$\times 10^{-4}$ &   0   &   6.9$\times 10^{-5}$ &   0   &   2.6$\times 10^{-4}$ \\
    &       &       &   (4.3$\times 10^{-4}$)   &       &   (7.0$\times 10^{-5}$)   &       &   (2.6$\times 10^{-4}$)   \\  \hline
    Be  &   300 &   0   &   0   &   0   &   0   &   0   &   $\Delta E<0$    \\
    &   650 &   0   &   0   &   0   &   0   &   0   &   $\Delta E<0$    \\
    & 1000  &   0   &   0   &   0   &   0   &   0   &   $\Delta E<0$    \\  \hline
Na  &   300 &   0   &   0   &   0   &   0   &   0   &   2.1$\times 10^{-4}$ \\
    &       &       &       &       &       &       &   (2.1$\times 10^{-4}$)   \\
    &   650 &   0   &   0   &   0   &   0   &   0   &   4.4$\times 10^{-3}$ \\
    &       &       &       &       &       &       &   (4.2$\times 10^{-3}$)   \\
    & 1000  &   0   &   0   &   0   &   1.8$\times 10^{-5}$ &   0   &   1.0$\times 10^{-2}$ \\
    &       &       &       &       &   (1.8$\times 10^{-5}$)   &       &   (0.96$\times 10^{-2}$)  \\  \hline
K   &   300 &   0   &   0   &   0   &   0   &   0   &   0   \\
    &   650 &   0   &   0   &   0   &   0   &   0   &   0   \\
    & 1000  &   0   &   0   &   0   &   0   &   0   &   3.6$\times 10^{-6}$ \\  \hline
Ca  &   300 &   0   &   0   &   0   &   0   &   0   &   0   \\
    &   650 &   0   &   0   &   0   &   0   &   0   &   1.1$\times 10^{-4}$ \\
    &       &       &       &       &       &       &   (1.0$\times 10^{-4}$)   \\
    &   1000    &   0   &   0   &   0   &   0   &   0   &   1.0$\times 10^{-3}$ \\
    &       &       &       &       &       &       &   (0.96$\times 10^{-3}$)  \\  \hline
Rb  &   300 &   0   &   0   &   0   &   0   &   0   &   0   \\
    &   650 &   0   &   0   &   0   &   0   &   0   &   0   \\
    &  1000 &   0   &   0   &   0   &   0   &   0   &   0   \\  \hline
Sr  &   300 &   0   &   0   &   0   &   0   &   0   &   0   \\
    &   650 &   0   &   0   &   0   &   0   &   0   &   0   \\
    & 1000  &   0   &   0   &   0   &   0   &   0   &   0   \\  \hline
Cs  &   300 &   0   &   0   &   0   &   0   &   0   &   0   \\
    &   650 &   0   &   0   &   0   &   0   &   0   &   0   \\
    &   1000    &   0   &   0   &   0   &   0   &   0   &   0   \\  \hline
Ba  &   300 &   0   &   0   &   0   &   0   &   0   &   0   \\
    &   650 &   0   &   0   &   0   &   0   &   0   &   0   \\
    &   1000    &   0   &   0   &   0   &   0   &   0   &   0   \\
\hline\hline \end{tabular} \end{table}

The calculated negligible solubilities of Be, Na, Ca in MgB$_2$
agree with previous experimental
observations\cite{Zhang02CaLi,Tamp03CaSr,Cheng03Ca,Toul02NaCa,Agost07Na,Feln01Be}.
We did not find any sign of nonvanishing solubilities of Rb, Cs and
Ba in the MgB$_2$ alloy even at high annealing temperature in
disagreement with the reported experimental data of Ref.
\onlinecite{Palnich07RbCsBa}. However, our negligible Rb and Cs
solubility in bulk MgB$_2$ are in agreement with the experimental
conclusion made in Ref. \onlinecite{Singh08RbCs} where the author
suggest that Rb and Cs dopants most likely segregate in grain
boundaries.

The obtained low solubilities of Li and Sr in MgB$_2$ differ from
experimental vales (solubility up to $30\%$ in Refs.
\onlinecite{Zhao01Li,Owens01Li,Cimb02Li,Zhang02CaLi,Tamp03CaSr,Karp08Li}).
This discrepancy can be attributed to segregation of Li and Sr in
the grain boundaries as was concluded for Rb and Cs in Ref.
\onlinecite{Singh08RbCs}. In particular, the value of
$E_\sol^{(s)}$=3.8 eV for Sr is too large for non-negligiblle bulk
solubility. For Li in MgB$_2$ the calculated formation energy
$E_\sol^{(s)}$=0.57 eV is not very large and it is the smallest
among the investigated solutes in MgB$_2$.
In Ref. \onlinecite{BernMass06}, qualitative conclusions about high
solubility of Li and low solubility of Na in MgB$_2$ were made based
on a large difference between the corresponding formation energies.
Our numerical results demonstrate that, despite much lower
formation energy of Li in comparison with Na in MgB$_2$, the
Li-solubility is still very low (Sec. \ref{SEC:FormEn.results}).

It should be emphasized that our model is in thermodynamical equilibrium which can be
difficult to reach at low temperatures and the experimental
equilibrium solubility tends usually to be overestimated. In fact,
the formation of metastable and/or unstable states which are
subsequently frozen at low temperatures, can make solubility
measurements very challenging. In such scenarios, the measured
solubility may correspond to spinodal concentration rather then
actual binodal concentration or simply characterize the frozen out
of equilibrium solubility remaining from the initial specimen
preparation at higher temperature. Besides, the segregation of
defects into grain-boundaries, especially in multicrystalline
samples prepared through non optimal cooling dramatically affect the
amount of frozen defects and solutes.

Although in our study we did not perform an extensive
search over the configurational space of ternary alloy, if a new
ternary phases were present near MgB$_2$, the
solute atoms would concentrate and nucleate such phases,
and this may be misinterpreted as a high solubility in MgB$_2$ phase.
Furthermore, a new equilibrium ternary phase would result in an increase
of impurities formation energies and, correspondingly, in a decrease of solubility.
Numerical approximations in the first-principles are not expected to
affect the values of solubility: i.e. tipically for $\delta E_\sol\sim 30$ meV/atom,
$\delta x_\A^{(1)} \sim 2 \times10^{-4}$.
Lattice vibrations and solute-solute interactions are neglected
because they are not expected to play important roles at low
temperatures and low solute concentrations.

\section{Conclusions} \label{SEC:conclusions}

In the present paper, we present an approach to study the
solubilities in ternary alloys. The advantage of the approach is in
taking into account all known ternary ground states rather than just
pure solids. Based on the approach, we propose an analytical
low-solubility approximation that can be used for high-throughput
calculations of solubilities in alloys.

Combining the developed approach with first principle calculations,
we have determined the formation energies and solubilities of alkali (Li,
Na, K, Rb, Cs) and alkaline earth (Be, Ca, Sr, Ba) metals in the Mg-B system.
It is found that the considered metals have low solubilities in the
boron-rich Mg-B alloy.
Substitutional Na, Ca, and Li experience the the largest solubilities, with Na in MgB$_7$
reaching 0.5-1\% at $T=650-1000$ K.
All the considered interstitial scenarios leed to negligible solubilities.
The solubility of Be in MgB$_7$ can not be determined with our model
because the corresponding low-solubility formation energy is negative implying
that the existing ground states list must be augmented through a more extensive search
over the configurational space.

We also present a high-throughput search of ground states
in binary Mg-B, Mg-$A$, and B-$A$ alloys ($A$=Li, Be, Na, Mg, K, Ca, Rb, Sr, Cs, Ba).
Ternary phase diagrams Mg-B-$A$ are constructed based on of the determined phases.
Sr$_{9}$Mg$_{38}$ is not an equilibrium ground state despite
its high temperature validations. Two new ground states CaB$_{4}$ and RbB$_{4}$ are found.

\section*{Acknowledgements}
We acknowledge Mike Mehl, Igor Mazin, and Wahyu Setyawan for fruitful
discussions. This research was supported by ONR
(Grants No. N00014-07-1-0878 and N00014-07-1-1085) and NSF (Grant No. DMR-0639822)
We thank the Teragrid Partnership (Texas Advanced Computing Center, TACC) for computational support (MCA-07S005).


\begin{thebibliography}{*}
\bibitem{note1}
  The bar at $\omu_s$ indicates that the quantity is not a
  chemical potential itself, as the number of particle is conserved.
\bibitem{note4}
  The approximation implies that a true-ternary ground state does not exist or, if it exist, it is in a
  region of the phase diagram not relevant for solubility of $A$ in MgB$_2$, MgB$_4$ and MgB$_7$.
\bibitem{note5}
  As shown in Sec. \ref{SEC:solubility.theory},
  the low-solubility formation energy
  $E_\sol^{(s)}$(Na)=1.53 eV rather than ``true'' formation energy
  $E_\texttt{\footnotesize Na}^{(s)}=1.67$ is more accurate for corresponding
  solubility characterization when the system is miscible.
  The 0.14 eV difference between $E_\texttt{\footnotesize Na}^{(s)}$
  and $E_\sol^{(s)}$(Na) suggests that the effect of other ground
  state (MgB$_4$, see Fig. \ref{Fig:PhDi-31}) on the formation energy
  and subsequently on the solubility is not negligible.
\end{thebibliography}
\end{document}